\documentclass[onecolumn]{emulateapj}

\shorttitle{Multi-layered configurations in differentially-rotational equilibrium}
\shortauthors{Kiuchi et al.}

\begin{document}


\title{Multi-layered configurations in differentially-rotational equilibrium}


\author{Kenta Kiuchi\altaffilmark{1}}
\affil{Department of Physics, Waseda University, 3-4-1 Okubo,
 Shinjuku-ku, Tokyo 169-8555, Japan~}

\author{Hiroki Nagakura\altaffilmark{2}}
\affil{Department of Physics, Waseda University, 3-4-1 Okubo,
 Shinjuku-ku, Tokyo 169-8555, Japan~}


\author{Shoichi Yamada\altaffilmark{3}}
\affil{Department of Physics, Waseda University, 3-4-1 Okubo,
 Shinjuku-ku, Tokyo 169-8555, Japan~}


\altaffiltext{1}{kiuchi@gravity.phys.waseda.ac.jp}
\altaffiltext{2}{hiroki@heap.phys.waseda.ac.jp}
\altaffiltext{3}{shoichi@waseda.jp}

\begin{abstract}
We present a new formula to numerically construct configurations in rotational equilibrium, which consist of multiple layers. 
Each layer rotates uniformly or differentially according to cylindrical rotation-laws that are different from layer to layer. Assuming  
a different barotropic equation of state (EOS) for each layer, we solve the Bernoulli equation in each layer separately and 
combine the solutions by imposing continuity of the pressure at each boundary of the layers. It is confirmed that a single continuous 
barotropic EOS is incompatible with the junction condition. Identifying appropriate variables to be solved, we construct a 
convergent iteration scheme. For demonstration, we obtain two-layered configurations, each layer of which rotates rapidly with 
either an "$\Omega$-constant law" or a "$j$-constant law" or a "$v$-constant  law". Other rotation laws and/or a larger 
number of layers can be treated similarly. We hope that this formula will be useful 
in studying the stellar evolution in multi-dimension with the non-spherical configuration induced by rotation being fully taken into account.
\end{abstract}


\section{Introduction}\label{sec:Intro}

It is well known that stars are generically rotating on the main sequence and, in particular, massive stars are rapid rotators~\citep{Fukuda:1982,Tassoul:2000}. 
Although the distribution of angular momentum in the stellar interior is poorly known except for the sun, it is expected that 
the inner portion is rotating more rapidly than the outer part as the star evolves and the central part of the star contracts.
In fact some recent theoretical studies on the evolution of rotating stars have demonstrated that massive stars in their late
evolutionary phases develop a central core that is rotating more rapidly than the outer envelopes~\citep{Heger:2000ud,Hirschi:2004ks,
Limongi:2000km}. 
Hence the differential rotation is supposed to exist quite commonly in the stellar interior especially at the advanced stages of evolution.

The above-mentioned works on the evolution of rotating stars ignore non-spherical deformations of stars and the angle-averaged centrifugal
force is added as a correction to the spherical models. This is not justified, however, if the star is rotating rapidly and the physical conditions
on the rotation axis and on the equatorial plane are substantially different. Then the rotational equilibrium should be properly taken into
account. This will be particularly important for the investigation of the progenitors of gamma ray bursts, since they are supposed to be driven
by the gravitational collapse of very rapidly rotating massive stars~\citep{Woosley:2005gy}.   

Over the years substantial effort has been made to numerically obtain configurations in rotational equilibrium in various contexts. 
Beginning with the pioneering works by \citet{Ostriker:1968}, a robust iterating formula was developed by~\citet{Hachisu:1986} and was extended to 
general relativistic and/or magnetized stars~\citep{Bocquet:1995,Cook:1993qj,Kiuchi:2008ch,Komatsu:1989,Tomimura:2005}. 
The polytropic equation of state (EOS) was replaced by more generic ones~\citep{Cook:1993qr,Kiuchi:2007pa,Kiuchi:2009zt}. 
One of the limitations of these studies that hamper the application to the study on the evolution of rotating stars, putting aside the treatment 
of convections and meridional circulations, is the assumption that the whole star is rotating cylindrically, that is, the angular velocity is 
constant on each of the concentric cylinders and, as a result, the EOS is barotropic, that is, the pressure is a function of density alone~\citep{Tassoul:2000}. 
On the contrary, the theoretical studies based on the spherical models indicate clearly that the stellar core and envelopes composed of 
different elements rotate rather independently of each other, since the steep gradient of mean molecular weight tends to suppress the 
transport of angular momentum beyond the boundary of elements. If one attempts to employ the rotational equilibrium configurations in 
the study of the post-main-sequence evolutions of massive stars, therefore, it is almost mandatory to treat multiple layers that obey 
different rotation-laws. 

Motivated by these facts, we present a new formulation to numerically construct non-relativistic configurations in rotational equilibrium, which consist of 
multiple layers. We assume that each layer rotates still cylindrically but the rotation-law, namely the angular velocity as a function of the 
distance from the rotation axis, can be different from layer to layer. This assumption allows us to employ the conventional formula based on 
the Bernoulli equation in each layer, a big advantage over the original partial differential equations. The EOS should be barotropic
in each layer accordingly. We then introduce a junction condition, that is, the continuity of pressure at the boundary of the layers to combine 
them in such a way that the whole star is in rotational equilibrium. In this paper, mainly for demonstration purposes, we obtain two-layered configurations 
with each layer having different polytropic EOS's for three representative rotation-laws:(1) $\Omega$-constant law (rigid rotation), 
(2) $j$-constant law and (3) $v$-constant law (see the next section for the exact definitions of these rotation laws).
In principle, there is no problem in treating different rotation-laws and/or a larger number of layers. 

The paper is organized as follows. In Sec.~\ref{sec:basic}, we describe the new formulation to obtain rotational equilibrium 
configurations with multiple layers, explaining numerical issues in detail. Section~\ref{sec:res} is devoted to the demonstration of some model 
computations. In Sec.~\ref{sec:sum}, we give some discussions and summarize the paper. 

\section{Basic equations \& Iteration Scheme}\label{sec:basic}
\subsection{Formulations}\label{sec:bs}
\begin{figure*}
  \begin{center}
  \vspace*{40pt}
    \begin{tabular}{c}
      \begin{minipage}{0.5\hsize}
      \includegraphics[width=9.5cm]{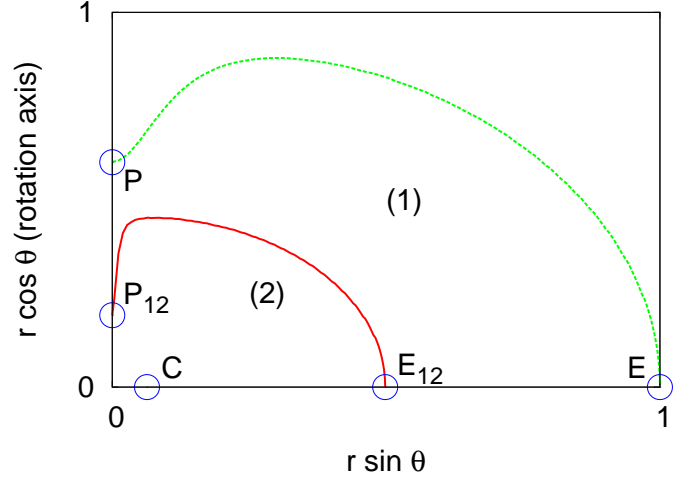}
      \end{minipage}
    \end{tabular}
    \caption{\label{fig:multi}
    Schematic picture of the two-layered configuration in rotational equilibrium 
    in the meridional section. The solid and dashed curves represent the layer boundary and 
    stellar surface, respectively. The circles attached by letters $C$, $E_{12}$, $E$, $P_{12}$ and $P$ 
    correspond to the point of density maximum, layer boundary and surface on equator and rotation axis,  
    respectively. Each layer is referred to by the number 1 or 2. See the text for more details.
    }
  \end{center}
\end{figure*}

In this paper the configurations in rotational equilibrium are assumed to be axisymmetric and steady with dissipative processes being neglected. 
Possible meridian flows are also ignored. Then the basic equations are 
\begin{eqnarray}
\label{eq:hydro2}
\frac{1}{\rho(r,\theta)}\nabla_i p(r,\theta) & = & -\nabla_i \phi_g(r,\theta) 
+ r\sin\theta \, \Omega^2(r,\theta) \nabla_i(r\sin\theta),  \\
\label{eq:gravity}
\nabla^{2}\phi_g(r,\theta) & = & 4\pi G \rho(r,\theta),
\end{eqnarray}
where the spherical coordinates are used and the subscript $i$ refers to each component. The gravitational potential, mass density, pressure, 
angular velocity and gravitational constant are denoted by $\phi_g$, $\rho$, $p$, $\Omega $ and $G$, respectively. For a barotropic EOS, i.e., $p=p(\rho)$, which we assume
for each layer throughout this paper, the left hand side of Eq.~(\ref{eq:hydro2}) can be integrated. The second term on the right hand side (RHS) of Eq.~(\ref{eq:hydro2}),
on the other hand, can be also integrated if the angular velocity is a function of the cylindrical radius, i.e., $\Omega=\Omega(r\sin\theta)$. 
As representative cases, the following three rotation laws are chosen in this paper although there is no limitation in principle:
\begin{eqnarray}
&&\Omega(r\sin\theta)=\Omega_0 \quad \text{(rigid rotation)},\label{eq:omeg}\\
&&\Omega(r\sin\theta)=\frac{j_0}{r^2\sin^2\theta+A^2} \quad \text{(constant specific angular momentum in the limit of $A \rightarrow 0$)},\label{eq:jconst}\\
&&\Omega(r\sin\theta)=\frac{v_0}{(r^2\sin^2\theta+A^2)^{1/2}} \quad \text{(constant rotational velocity in the limit of $A \rightarrow 0$)}.\label{eq:vconst}
\end{eqnarray}
In these expressions, $\Omega_0$, $j_0$ and $v_0$ are constants that specify how fast the rotation is for each rotation law whereas $A$ 
is a constant that gives a radius of the cylinder, inside which the rotation is almost rigid. The second and third rotation laws are referred to 
as the $j$-constant and $v$-constant laws, respectively, according to their limits of $A \rightarrow 0$. The first rotation law is also called $\Omega$-constant 
law. As already mentioned, the constants $\Omega_0$, $j_0$, $v_0$ and $A$ can be different from layer to layer for multi-layered 
configurations. The values of these constants in the $i$-th layer (see Fig.~\ref{fig:multi}.) are represented by the subscript $(i)$ in the following. 

The integration of Eq.~(\ref{eq:hydro2}) in each layer gives the so-called Bernoulli equation, 
\begin{eqnarray}
H_{(i)}(\rho(r,\theta))=-\phi_g(r,\theta) + h_{0(i)}^2 \phi_{\rm rot}(r\sin\theta:A_{(i)})+c_{(i)}, \label{eq:Ber}
\end{eqnarray}
where $H_{(i)}$ is a specific enthalpy defined by $\int dP/\rho$ as a function of the mass density alone and $c_{(i)}$ is called 
a Bernoulli constant. The second term on RHS corresponds to the rotational potential and is given for each rotation law as 
\begin{eqnarray}
&&h_{0(i)}=\Omega_{0(i)}, \quad \phi_{\rm rot}(r\sin\theta) = \frac{1}{2}r^2\sin^2\theta \label{eq:rot0} \quad \text{(rigid rotation)},\\
&&h_{0(i)}=j_{0(i)}, \quad \phi_{\rm rot}(r\sin\theta:A_{(i)})=-\frac{1}{2(r^2\sin^2\theta+A^2_{(i)})} \label{eq:rot1} \quad \text{(j-constant law)},\\
&&h_{0(i)}=v_{0(i)}, \quad \phi_{\rm rot}(r\sin\theta:A_{(i)})=\frac{1}{2}\ln(r^2\sin^2\theta+A^2_{(i)}) \label{eq:rot2} \quad \text{(v-constant law)},
\end{eqnarray}
where the amplitude, $h_{0(i)}$, is expressed separately for later convenience. Note that the Bernoulli constants can be different from
layer to layer. It is well known that the barotropic condition is equivalent to the requirement that the angular velocity be a function of the 
cylindrical radius alone, i.e., $\Omega=\Omega(r\sin\theta)$~\citep{Tassoul:2000}. 
In this article, we assume a different rotation law for each layer, which means that the angular velocity is not a function of the 
cylindrical radius alone even if the rotation law is cylindrical in each layer. As a result, the EOS cannot be a single continuous 
barotropic one. In fact, the angular velocity is discontinuous across the layer boundary, which leads in general to the discontinuity in density 
as shown later. On the other hand, the pressure is continuous at the layer boundary, which can be understood as follows: Multiplied by the density,
Eq.~(\ref{eq:hydro2}) is written at the layer boundary as  
\begin{eqnarray}
\nabla_i p =-\rho\nabla_i \phi_g
+ \rho r\sin\theta \, \Omega^2_{(i)} \nabla_i(r\sin\theta).\label{eq:hydro3}
\end{eqnarray}
The right hand side (RHS) of this equation contains step functions, that is, the density and angular velocity. Note that the gravitational potential, which is
obtained by the integration of the density, is continuous. Then the pressure is also continuous because otherwise the left hand side of Eq.~(\ref{eq:hydro3}) would 
give a delta function. This in turn leads to the conclusion that a single continuous barotropic EOS cannot be applied across the layer boundary, since 
the pressure could not be continuous for the discontinuous density for such an EOS. Therefore, EOS's that are barotropic in each layer but different 
from layer to layer are required. As the simplest example, we employ in this article polytropic EOS's with a different polytropic constant and/or index in each layer. 
The essential points of our formula are summarized as follows: (1) the rotational equilibrium is locally ensured by the Bernoulli equation in each layer and 
(2) the layer boundary is the location where the layers are joined so that the pressure should become continuous. 

\subsection{HSCF scheme for single-layered configurations}\label{sec:hscf}

The problem is now reduced to the solution of Eqs.~(\ref{eq:Ber}) and (\ref{eq:gravity}) and the search of the location where the pressures of the different layers 
coincide. Before discussing multi-layered configurations, we briefly review the Hachisu Self-Consistent Field (HSCF) 
scheme~\citep{Hachisu:1986}, which is known to be a very robust 
algorithm to solve iteratively Eqs.~(\ref{eq:Ber}) and (\ref{eq:gravity}) for single-layered configurations
in rotational equilibrium and on which our formula is based.  
In this scheme, we first introduce the following non-dimensional variables: 
\begin{eqnarray}
&&\hat{\rho}=\rho / \rho_{\rm max}, \label{eq:nor1}\\
&&\hat{r}= r / r_e, \label{eq:defofbeta}\\
&&\hat{h}_0^2 \hat{\phi}_{\rm rot} = h_0^2\phi_{\rm rot}/4\pi G\rho_{\rm max} r_e^2, \\
&&\hat{c}= c / 4\pi{\rm G}\rho_{\rm max} r_e^2,  \\
&&\hat{\phi}_g= \phi_g / 4\pi{\rm G}\rho_{\rm max} r_e^2,\\
&&\hat{p} = p/\rho_{\rm max} \, c^2,\\
&&\hat{H} = H/c^2,\label{eq:nor2}
\end{eqnarray}
where $\rho_{\rm max}$, $p_{\rm max}$ and $c$ are the maximum density, pressure and speed of light, respectively, 
and the subscript $i$ is dropped in $h_0$, $A$, $c$ and $H$. The radius is normalized by the 
equatorial radius of the equilibrium configuration, $r_e$, which is unknown a priori and is expressed 
as $r_{e}=\displaystyle{\sqrt{\frac{1}{\beta} \frac{p_{\rm max}}{4\pi {\rm G}\rho_{\rm max}^2}}}$ by the introduction of
a new variable $\beta$. Then Eqs.~(\ref{eq:Ber}) and (\ref{eq:gravity}) are reduced to 
\begin{eqnarray}
\label{eq:Ber2}
&&\frac{\beta}{\hat{p}_{\rm max}} \hat{H} (\hat{\rho}(\hat{r},\theta))
= - \hat{\phi}_g (\hat{r},\theta)+ \hat{h}_0^2 \hat{\phi}_{\rm rot} (\hat{r}\sin\theta:A)+ \hat{c}, \\
&&\hat{\nabla}^2 \hat{\phi}_g(\hat{r},\theta) = \hat{\rho}(\hat{r},\theta)\label{eq:pois2}.
\end{eqnarray}
In Eqs.~(\ref{eq:Ber2}) and (\ref{eq:pois2}), we have two unknown functions $\hat{\phi}_g(\hat{r},\theta)$ and $\hat{\rho}(\hat{r},\theta)$ 
and three constants $\beta$, $\hat{h}_0$, and $\hat{c}$ once the EOS and rotation law (\ref{eq:rot0}-\ref{eq:rot2}) are 
specified. It is noted that the specific enthalpy $H$ is a function of the density alone because of the barotropic 
condition. In the HSCF method, we give $\rho_{\rm max}$, the equatorial radius $\hat{r}_e$ and polar radius $\hat{r}_p$
instead of $\beta$, $\hat{h}_0,$ and $\hat{c}$ to specify the model and the latter three are treated as unknown variables to
be solved. Note that $\hat{r}_e$ is unity by the definition of $\beta$ (see Eq.~(\ref{eq:defofbeta})). 
This choice of variables is essential for the HSCF scheme. Indeed other choices such as $\rho_{\rm max}$ and $\hat{h}_0$ 
(and $\hat{r}_e=1$) fail to obtain convergence in the iteration (see below) more often than not.  

The two unknown functions $\hat{\phi}_g(\hat{r},\theta)$ and $\hat{\rho}(\hat{r},\theta)$ 
and three unknown constants $\beta$, $\hat{h}_0$, and $\hat{c}$ are obtained iteratively in the HSCF method as follows. 
First we give a trial density distribution $\hat{\rho}$ and solve Eq.~(\ref{eq:pois2}) to obtain $\hat{\phi}_g$. As a second step, 
Eq.~(\ref{eq:Ber2}) is evaluated at the following points: the stellar surfaces on the equator and on the rotation axis and the point of 
the maximum density, which are denoted as $E$, $P$ and $C$, respectively (see Fig.~\ref{fig:multi}).
\begin{eqnarray}
(E)&& \quad 0= - \hat{\phi}_g (1,\pi/2)+ \hat{h}_0^2 \hat{\phi}_{\rm rot} (1:A)+ \hat{c}, \label{eq:Hac0}\\
(P)&& \quad 0= - \hat{\phi}_g (\hat{r}_p,0)+ \hat{h}_0^2 \hat{\phi}_{\rm rot} (0:A)+ \hat{c}, \label{eq:Hac1}\\
(C)&& \quad \frac{\beta}{\hat{p}_{\rm max}} \hat{H}(\hat{\rho}_{\rm max})
= - \hat{\phi}_g (\hat{r}_C,\pi/2)+ \hat{h}_0^2 \hat{\phi}_{\rm rot} (\hat{r}_C:A)+ \hat{c}, \label{eq:Hac2}
\end{eqnarray}
where we made use of the fact that the enthalpy vanishes on the stellar surface. Note that the radius, $\hat{r}_C$, of point C 
is not known a priori. For the rigid rotation $\hat{\phi}_{\rm rot}$ is independent of $A$. We solve Eqs.~(\ref{eq:Hac0}) and (\ref{eq:Hac1}) 
with respect to $\hat{c}$ and $\hat{h}_0$ for $\hat{\phi}_g(\hat{r},\theta)$ obtained in the first step. With these values of $\hat{h}_0$ and $\hat{c}$, 
we then search for the location where the RHS of Eq.~(\ref{eq:Hac2}) takes the maximum value. The maximum of the RHS of Eq.~(\ref{eq:Hac2})
thus obtained gives $\beta$ in turn. We are now in a position to update $\hat{\rho}(\hat{r},\theta)$, solving the Bernoulli equation (\ref{eq:Ber2}) for 
$\hat{\phi}_g(\hat{r},\theta),\beta,\hat{h}_{0},$ and $\hat{c}$ obtained so far. We then repeat the procedure until a sufficient convergence is achieved. 

\subsection{Extension to multi-layered configurations}\label{sec:extmul}
\subsubsection{choice of variables}\label{sec:chovar}
We move on to the multi-layered case. Although, for simplicity, we consider only two-layered structures in the following, the extension to 
configurations with a larger number of layers is straightforward. Then we have again two unknown functions $\hat{\phi}_g(\hat{r},\theta)$ and 
$\hat{\rho}(\hat{r},\theta)$ and this time five constants $\beta$, $\hat{h}_{0(i)}$, and $\hat{c}_{(i)}$ in Eqs.~(\ref{eq:Ber}) and (\ref{eq:gravity}). 
Another important function to be determined is the layer boundary expressed by a function $\hat{r}_{b}(\theta)$. Just as in the single layer case,
the choice of variables and the iteration scheme are critically important to make the scheme convergent. We first write down the equations 
employed to obtain the unknown constants, which correspond to Eqs.~(\ref{eq:Hac0})-(\ref{eq:Hac2}) for the single-layered case.
As explained above, the pressure should be continuous across the layer boundary. Employing this condition on the equator 
(point $E_{12}$ in Figure~\ref{fig:multi}) and on the rotation axis (point $P_{12}$ in the same figure), we write down  
the Bernoulli equation for both sides of the layer boundary at these points:
\begin{eqnarray}
\text{(layer 1 side at $E_{12}$)}&&\quad \frac{\beta}{p_{\rm max}} H_{(1)}(p(r_{E_{12}},\pi/2))
= - \phi_g (r_{E_{12}},\pi/2)+ h_{0(1)}^2\phi_{\rm rot} (r_{E_{12}}:A_{(1)})+ c_{(1)}
,\label{eq:cond2}\\
\text{(layer 2 side at $E_{12}$)}&&\quad \frac{\beta}{p_{\rm max}} H_{(2)}(p(r_{E_{12}},\pi/2))
= - \phi_g (r_{E_{12}},\pi/2)+ h_{0(2)}^2\phi_{\rm rot} (r_{E_{12}}:A_{(2)})+ c_{(2)}
,\label{eq:cond1}\\
\text{(layer 1 side at $P_{12}$)}&&\quad \frac{\beta}{p_{\rm max}} H_{(1)}(p(r_{P_{12}},0))
= - \phi_g (r_{P_{12}},0)+ h_{0(1)}^2\phi_{\rm rot} (0:A_{(1)})+ c_{(1)}
,\label{eq:cond4}\\
\text{(layer 2 side at $P_{12}$)}&&\quad \frac{\beta}{p_{\rm max}} H_{(2)}(p(r_{P_{12}},0))
= - \phi_g (r_{P_{12}},0)+ h_{0(2)}^2\phi_{\rm rot} (0:A_{(2)})+ c_{(2)}
,\label{eq:cond3}
\end{eqnarray}
where we omit $\hat{ }$ over the normalized variables for notational simplicity and $r_{E_{12}}$, $r_{P_{12}}$ are the radii of points 
$E_{12}$ and $P_{12}$, respectively. Note that $H_{(1)}(p)\ne H_{(2)}(p)$ because the EOS's are different from layer to layer. 
We employ the previous conditions at points $E$, $P$ and $C$, which are written as 
\begin{eqnarray}
(E) &&\quad 0= - {\phi}_g (1,\pi/2)+ {h}_{0(1)}^2 {\phi}_{\rm rot} (1:A_{(1)})+ {c}_{(1)},\label{eq:cond5}\\
(P) &&\quad 0= - {\phi}_g ({r}_p,0)+ {h}_{0(1)}^2 {\phi}_{\rm rot} (0:A_{(1)})+ {c}_{(1)},\label{eq:cond6}\\
(C) && \quad \frac{\beta}{p_{\rm max}} H_{(i)}({\rho}_{\rm max})
= - {\phi}_g ({r}_C,\pi/2)+ {h}_{0(i)}^2 {\phi}_{\rm rot} ({r}_C:A_{(i)})+ {c}_{(i)}. \quad (i=1 \ {\rm or}\ 2)\label{eq:cond7}
\end{eqnarray}
Note that in Eq.~(\ref{eq:cond7}) we do not know a priori in which layer the maximum density point $C$ lies. 
Using Eqs.~(\ref{eq:cond2})-(\ref{eq:cond7}) we can determine for the given EOS's and rotation laws seven unknown constants out of  
$\beta$, $h_{0(1)}$, $h_{0(2)}$, $c_{(1)}$, $c_{(2)}$, $p_{E_{12}}$, $p_{P_{12}}$, $r_p$, $r_{E_{12}}$, $r_{P_{12}}$, in which $p_{E_{12}}\equiv p(r_{E_{12}},\pi/2)$ 
and $p_{P_{12}}\equiv p(r_{P_{12}},0)$. This implies that one can give three constants to specify the model. 
As argued  shortly, however, $p_{P_{12}}$ and $r_{P_{12}}$ as well as $p_{E_{12}}$ and $r_{E_{12}}$ can not be specified independently. 
This can be understood by considering a non-rotating but two-layered configuration. In this case one can construct an equilibrium configuration by  
integrating Eq.~(\ref{eq:hydro2}) radially from the center to $r_{P_{12}}$ with the use of the gravitational potential $\int^r_0 4\pi {r'}^2 \rho dr'/r^2$. 
Then it is obvious that $p_{P_{12}}$ depends on $r_{P_{12}}$ and vice versa (note that the maximum density is also fixed). 
Going back to the multi-layered rotational case, we have found that the combination of $r_p$, $r_{P_{12}}$ and $r_{E_{12}}$ is a good choice. 
The triplet of $r_p$, $p_{P_{12}},$ and $p_{E_{12}}$ can be an alternative. Note that the inclusion of $r_p$ seems to be mandatory as 
has been demonstrated by the HSCF scheme. To summarize, Eqs.~(\ref{eq:cond2})-(\ref{eq:cond7}) are used to obtain either $\beta$, $c_{(1)}$,
$c_{(2)}$, $h_{0(1)}$, $h_{0(2)}$, $p_{E_{12}}$ and $p_{P_{12}}$ for given $r_p$, $r_{E_{12}}$ and $r_{P_{12}}$ or 
$\beta$, $c_{(1)}$, $c_{(2)}$, $h_{0(1)}$, $h_{0(2)}$, $r_{E_{12}}$ and $r_{P_{12}}$ for given $r_p$, $p_{E_{12}}$ and $p_{P_{12}}$. 

\subsection{iteration scheme}\label{sec:iers}
Now we proceed to the iteration scheme proposed in this paper.  
After solving the Poisson equation for the trial density distribution,  
Eqs.~(\ref{eq:cond2})-(\ref{eq:cond7}) are solved for the variables chosen in the previous section. 
The procedure is divided into the following three steps:\\
(1) $h_{0(1)}$ and $c_{(1)}$ are calculated from Eqs.~(\ref{eq:cond5}) and (\ref{eq:cond6}) as
\begin{eqnarray}
&&h_{0(1)}^2 =  \frac{\phi_g(1,\pi/2)-\phi_g(r_p,0)}{\phi_{\rm rot}(1:A_{(1)})-\phi_{\rm rot}(0:A_{(1)})},\label{eq:h01}\\
&&c_{(1)}=\frac{-\phi_g(1,\pi/2)\phi_{\rm rot}(0:A_{(1)})+\phi_g(r_p,0)\phi_{\rm rot}(1:A_{(1)})}
{\phi_{\rm rot}(1:A_{(1)})-\phi_{\rm rot}(0:A_{(1)})}.
\end{eqnarray}
(2) For a trial value of $\beta$, $p_{E_{12}}$ and $p_{P_{12}}$ are obtained from 
Eqs.~(\ref{eq:cond2}) and (\ref{eq:cond4}) and, combined with Eqs.~(\ref{eq:cond1}) and (\ref{eq:cond4}), give $h_{0(2)}$ as 
\begin{eqnarray}
h_{0(2)}^2&=&\frac{\phi_{\rm rot}(r_{E_{12}}:A_{(1)})-\phi_{\rm rot}(0:A_{(1)})}
{\phi_{\rm rot}(r_{E_{12}}:A_{(2)})-\phi_{\rm rot}(0:A_{(2)})}h_{0(1)}^2\nonumber\\
&+&\frac{
 H_{(2)}(p_{E_{12}})-H_{(1)}(p_{E_{12}})
-H_{(2)}(p_{P_{12}})+H_{(1)}(p_{P_{12}})
}{\phi_{\rm rot}(r_{E_{12}}:A_{(2)})-\phi_{\rm rot}(0:A_{(2)})}
\frac{\beta}{p_{\rm max}}.\label{eq:h02}
\end{eqnarray}
Then $c_{(2)}$ is obtained from Eq.~(\ref{eq:cond1}) or (\ref{eq:cond3}).\\
(3) The point that gives the largest value to $-\phi_g(r,\pi/2) + h_{0(1)}^2 \phi_{\rm rot}(r:A_{(1)}) + c_{(1)}$
is searched in layer 1 and is referred to as point $C1$. The counter part in layer 2 is then looked for and is called point $C2$. 
The point $C$ is found by comparing $C1$ and $C2$ to $E_{12}$.
The maximum value thus obtained is divided by $H_{(i)}(\rho_{\rm max})/p_{\rm max}$ with the appropriate $i$ and the updated 
value of $\beta$ is obtained. The steps (2) and (3) are repeated until the value of $\beta$ converges at a sufficient level. 

\subsubsection{Layer boundary}\label{sec:laybou}
The final step of the iteration is the updates of the density distribution and the location of the layer boundary. This is 
accomplished as follows. Regarding the specific enthalpy as a function of pressure alone, we first solve the Bernoulli 
equation, Eq.~(\ref{eq:Ber}), for each layer in the absence of the other layer and the layer boundary as a result:
\begin{eqnarray}
&&p_{(1)}(r,\theta)=H^{-1}_{(1)}
\left(
(-\phi_g(r,\theta)+h_{0(1)}^2\phi_{\rm rot}(r\sin\theta:A_{(1)}) + c_{(1)})\frac{p_{\rm max}}{\beta}
\right),\\
&&p_{(2)}(r,\theta)=H^{-1}_{(2)}
\left(
(-\phi_g(r,\theta)+h_{0(2)}^2\phi_{\rm rot}(r\sin\theta:A_{(2)}) + c_{(2)})\frac{p_{\rm max}}{\beta}
\right),
\end{eqnarray}
where $H^{-1}_{(i)}$ is an inverse function of the specific enthalpy for layer $i$. Since the pressure should be continuous at the layer boundary 
as discussed in Sec.~\ref{sec:bs}, we look for a point on each radial ray (a line with $\theta=$ const.), at which $p_{(1)}$ and 
$p_{(2)}$ coincides with each other. This gives the updated layer boundary as $r=r_{b}(\theta)$. On the other hand, 
$p_{(1)}(r,\theta)$ and $p_{(2)}(r,\theta)$ obtained above give the updated density distribution for each layer. This closes the iteration
procedure. We go back to the Poisson equation and repeat all the steps until a sufficient level of convergence is reached.

\section{Result}\label{sec:res}

To demonstrate that the new formula described above really works, we will apply it to two-layered configurations 
for some representative rotation laws. For simplicity, we employ two polytropic EOS's, $p_{i}(\rho)=K_i\rho^{1+1/n_i}$, in which 
$K_i$ and $n_i$ are the polytropic constants and indices for layer $i$. We take rather arbitrarily 
$K_1=1.015\times 10^{15}$ and $K_2=5.14\times 10^{14}$ in cgs unit and $n_1=n_2=3$.
Note that the specific enthalpy is given as $H_{(i)}=(n_i+1)K_i \rho^{1/n_i}=n_i K_i^{n_i/n_i+1}p^{1/1+n_i}$ for layer $i$. 
Adopting $\rho_{\rm max}=5.629\times 10^9 {\rm g/cm^3}$ we obtain white dwarf-like configurations.

We work with the normalized variables (Eqs.~(\ref{eq:nor1})-(\ref{eq:nor2})) and the numerical domain covers 
a quadrant of the meridian section, $0 \leqq r \leqq 1$ and $0 \leqq \theta \leqq \pi/2$, under the assumption of 
equatorial symmetry. We typically deploy 1000 mesh points on the $r$-coordinate and 200 grid points 
on the $\theta$-coordinate to obtain an acceptable accuracy, which is  confirmed by the normalized virial 
equation~\citep{cow} defined as 
\begin{eqnarray}
VC =|2T+W+3U|/|W|, \label{eq:vir}
\end{eqnarray}
where $T,W,$ and $U$ are the rotational, gravitational and internal energies, respectively. The ratio should vanish
for exact solutions. Note finally that the Poisson equation is solved by the Green function method employed by~\cite{Hachisu:1986}. 
 
\subsection{Non-rotational case}
As a mandatory step, we first construct a non-rotational configuration with two layers according to the present formula and 
compare it with the solution obtained by the ordinary and much simpler method, that is, the radial integration. 
Setting $r_p=1$ and $r_{E_{12}}=r_{P_{12}}$, we obtain $h_{0(1)}=0$ from Eq.~(\ref{eq:h01}) and $p_{E_{12}}=p_{P_{12}}$ from  
Eqs.~(\ref{eq:cond2}) and (\ref{eq:cond4}) that are actually identical with each other. 
Then, Eq.~(\ref{eq:h02}) gives $h_{0(2)}=0$ because the first term vanishes owing to $h_{0(1)}=0$ 
and the second term is also zero because of $p_{E_{12}}=p_{P_{12}}$. As a consequence, Eq.~(\ref{eq:cond1}) becomes
identical with Eq.~(\ref{eq:cond3}). It should be noted that $c_{(1)}\ne c_{(2)}$ even in the non-rotational case, since the 
EOS's are different between two layers. 

Table~\ref{tab:test} summarizes the comparison of the non-rotating configuration obtained by the new formulation, which is 
referred to as "2D" in the table, with the solution of the one-dimensional hydrostatic equation obtained by the radial integration 
from center to surface, which is called "1D". The radius, mass and the Bernoulli constants in 2D, which characterize the equilibrium 
configuration, nicely agree with the 1D counter parts. Indeed, the relative errors in these quantities are less than 
one percent and the normalized virial is of the order of $10^{-5}$, both of which indicate that the present method can reproduce 
the non-rotational configuration. 
\begin{table}
\centering
\caption{\label{tab:test}
Comparison of the non-rotational two-layered configurations obtained by the ordinary method (1D) and the present formula (2D). 
The radius, mass in cgs unit and normalized virial are denoted by $R$, $M$, and $VC$, respectively.
For this model we adopt $r_{E_{12}}=r_{P_{12}}=1.718\times 10^{-2}$.}
\begin{tabular}{cccccc}
\hline\hline
   &$R~[{\rm cm}]$ & $M~[M_\odot]$ & $c_{(1)}$    & $c_{(2)}$    & $VC$\\
\hline
1D &$3.218$E+08    & $4.292$E+00   & $-3.622$E-03 & $-8.579$E-03 & --\\
2D &$3.224$E+08    & $4.291$E+00   & $-3.601$E-03 & $-8.533$E-03 & $2.815$E-05\\
\hline
\end{tabular}
\end{table}
\subsection{$\Omega$-constant case}
We now proceed to the rotational cases. In this section we deal with the simplest one, that is, the combination of 
two rigid rotations ($\Omega$-constant laws in Eq.~(\ref{eq:omeg})). We have constructed two configurations with 
$(r_p,r_{E_{12}},r_{P_{12}})=(0.9,0.5,0.5)$ and $(0.8,0.5,0.433)$. In Table~\ref{tab:twolay-R}, we show some key quantities
that characterize the configurations. As expected, the angular velocities are different between the layers. Note that 
in the present formula, the angular velocity is not specified but solved. It is $r_{E_{12}}$ and $r_{P_{12}}$ that dictate the 
angular velocities. If one wants to construct a configuration for particular angular velocities, another iteration is 
needed for shooting. Incidentally, we can impose in principle the combinations that satisfy $r_{E_{12}}\le r_{P_{12}}$, 
which is intuitively unlikely because rotations tend to flatten equilibrium configurations in general~\citep{Hachisu:1986}. 
In fact, we find that the solutions for such cases have a negative centrifugal force, which is, of course, unphysical. 
We thus confirm that the inner layer is still oblate in the multi-layered configurations.

\begin{figure*}
  \begin{center}
  \vspace*{40pt}
    \begin{tabular}{cc}
      \hspace{-2.0cm}
      \begin{minipage}{0.5\hsize}
      \includegraphics[width=12.0cm]{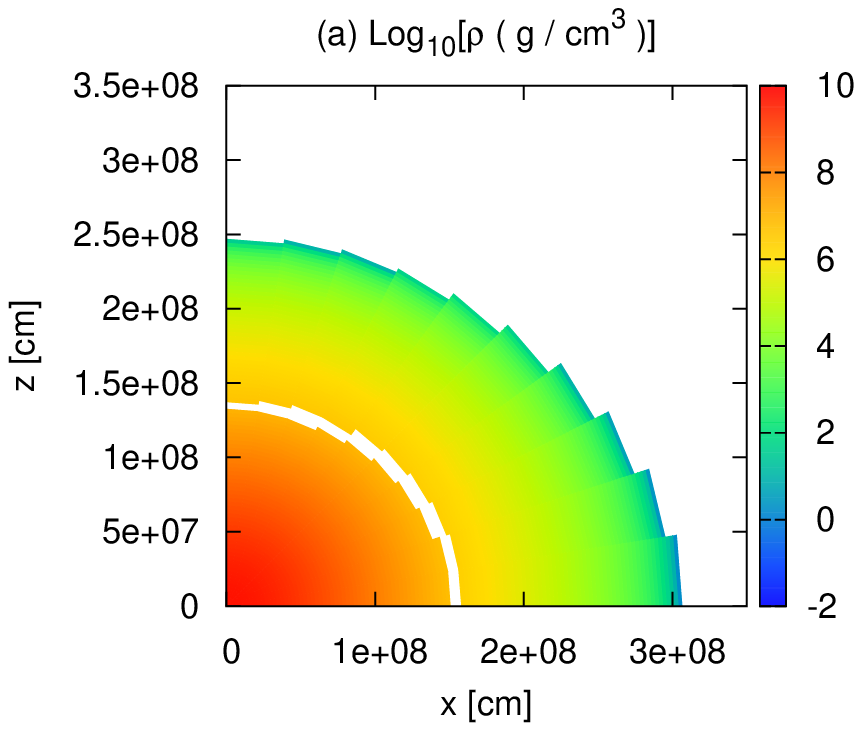}
      \end{minipage}
      \hspace{-1.0cm}
      \begin{minipage}{0.5\hsize}
      \includegraphics[width=12.0cm]{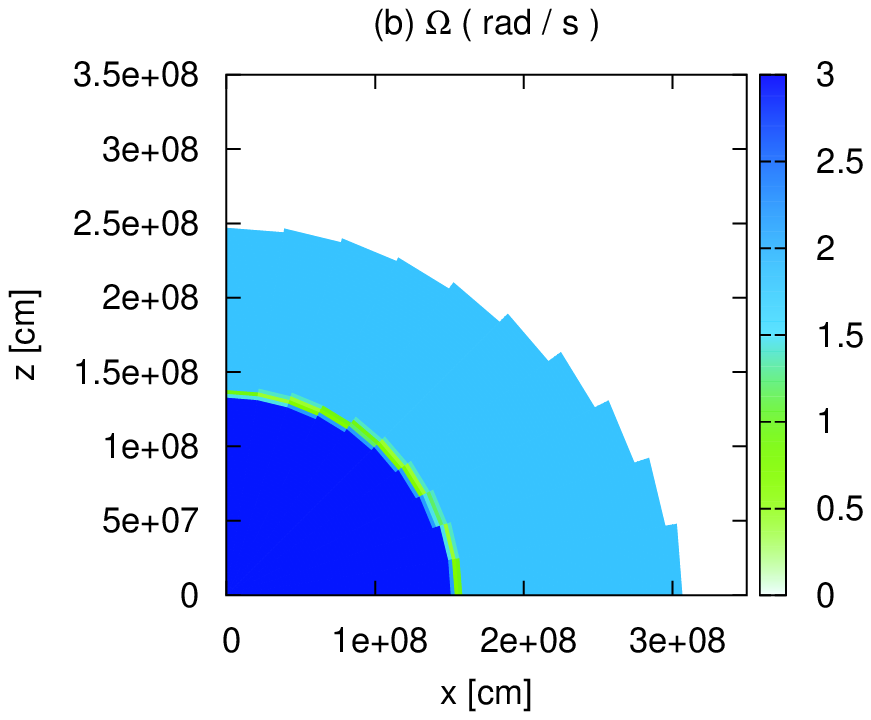}
      \end{minipage}
    \end{tabular}
    \caption{\label{fig:rigid}
    Contour plots in the meridional section of the density (left panel) and angular velocity (right panel) for 
    the $\Omega$-constant case. The contour levels are equally spaced in the logarithmic scale in panel~(a) 
    and in the linear scale in (b). The thick curves in the stellar interiors correspond to the layer boundary 
    in both panels. For this model $(r_p,r_{E_{12}},r_{P_{12}})=(0.8,0.5,0.433)$ are adopted.
    Each layer is rotating rigidly.}
  \end{center}
\end{figure*}

\begin{figure*}
  \begin{center}
  \vspace*{40pt}
    \begin{tabular}{cc}
      \hspace{-2.0cm}
      \begin{minipage}{0.5\hsize}
      \includegraphics[width=9.0cm]{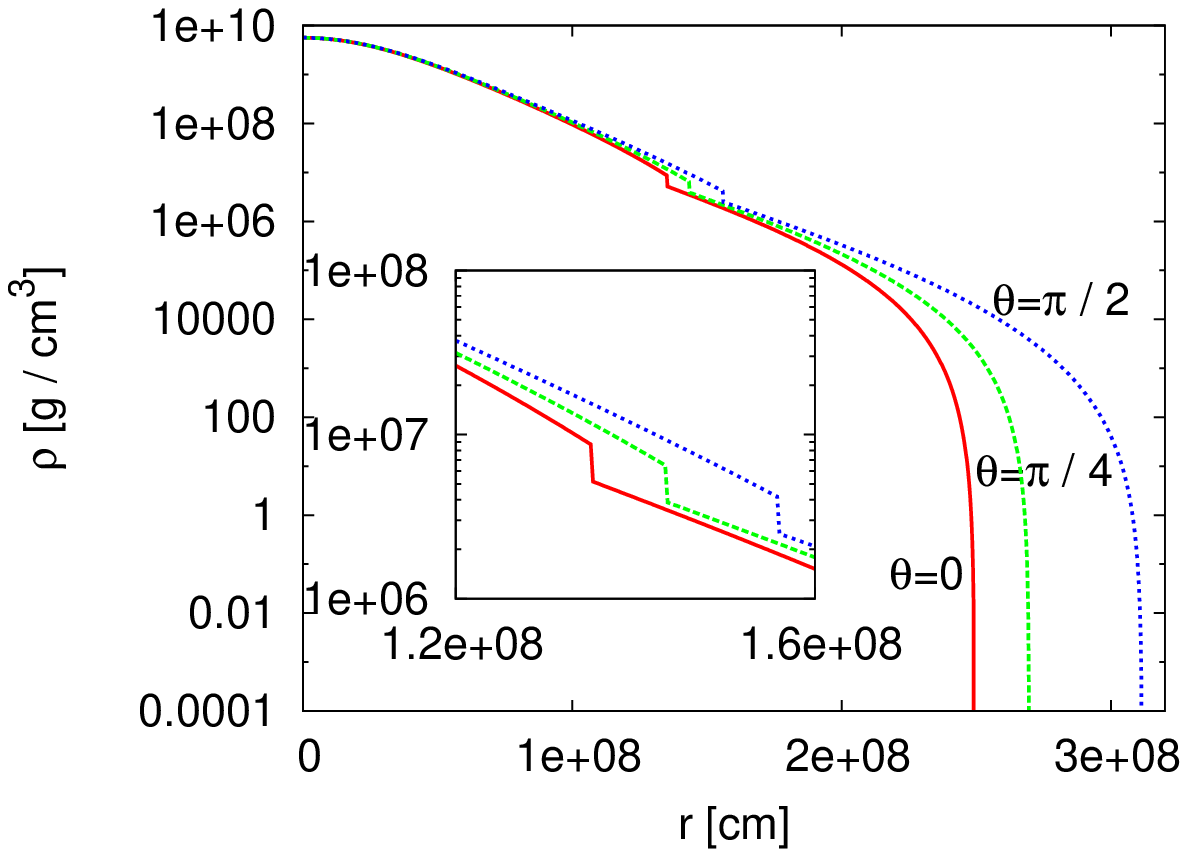}
      \end{minipage}
      \hspace{-0.5cm}
      \begin{minipage}{0.5\hsize}
      \includegraphics[width=9.0cm]{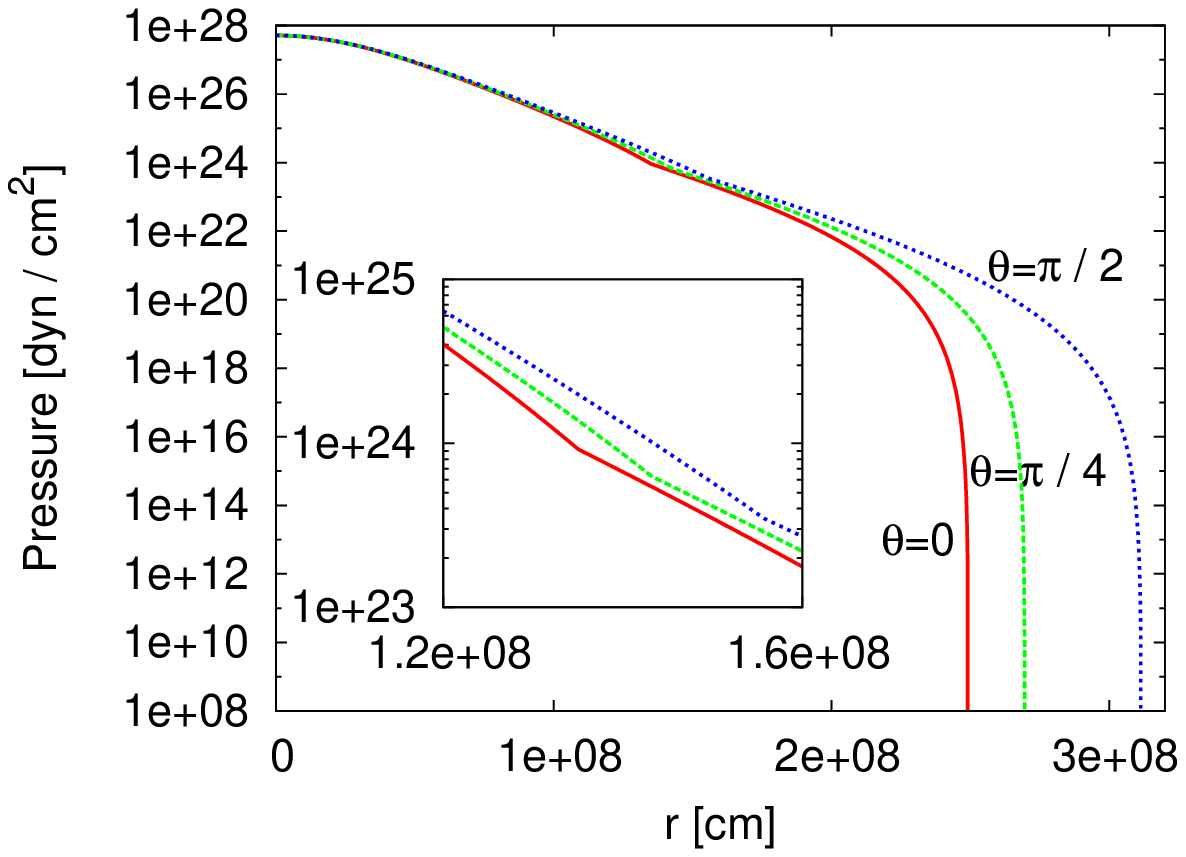}
      \end{minipage}
    \end{tabular}
    \caption{\label{fig:p} Density (left panel) and pressure (right panel) profiles 
    for the model in Fig.~\ref{fig:rigid}. 
    In both panels, the solid, dashed and dotted curves show the distributions along 
    the radial lines with $\theta=0$, $\pi/4$ and $\pi/2$, respectively. 
    The insets are the magnification of the vicinity of layer boundary. 
    }
  \end{center}
\end{figure*}

Figure~\ref{fig:rigid} displays the contour plots of the density and angular velocity in the meridian section for the 
model with $(r_p,r_{E_{12}},r_{P_{12}})=(0.8,0.5,0.433)$. The thick curve in the figure represents the layer boundary. 
It should be noted that the density is discontinuous at the layer boundary as mentioned earlier. This is generally the case.
On the other hand, the pressure should be continuous across the boundary. To check this, we show in Fig.~\ref{fig:p} 
the density and pressure profiles along the radial lines with $\theta=0$, $\pi/4$ and $\pi/2$. It is clear that 
the density is discontinuous at $1.2\times 10^{8}{\rm cm}\alt r \alt 1.6\times 10^{8}$ cm whereas the pressure profiles are 
continuous in the same region. From this we can understand again why different EOS's are needed for each layer. 

\begin{figure*}
  \begin{center}
  \vspace*{40pt}
    \begin{tabular}{c}
      \hspace{-2.0cm}
      \begin{minipage}{0.5\hsize}
      \includegraphics[width=9.0cm]{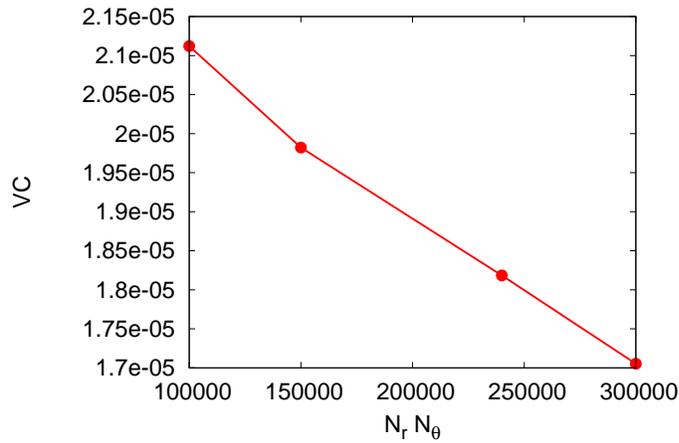}
      \end{minipage}
    \end{tabular}
    \caption{\label{fig:vir}
    Convergence test. The normalized virial is shown as a function of the number of grid points for the $\Omega$-constant model with 
    $(r_p,r_{E_{12}},r_{P_{12}})=(0.8,0.5,0.433)$. $N_r$ and $N_\theta$ denote the numbers of grid points on $r$- and $\theta$-coordinates, 
    respectively. 
    }
  \end{center}
\end{figure*}

The values of $VC$ in Table~\ref{tab:twolay-R} are of the same orders of magnitude, $\sim 10^{-5}$, as in the non-rotational case.
This implies that the configurations we have constructed by the new formula are in rotational equilibrium to the same accuracy
as the spherical configuration in the previous section is in hydrostatic equilibrium. Figure~\ref{fig:vir} plots the values of $VC$ as a function of the number of 
grid points. It is clearly demonstrated that the accuracy is increased as the resolution becomes better although the 
convergence is rather slow. We infer that this slow convergence is due to the discontinuity in the density distribution, 
which enters the virial equation through the gravitational binding energy $W$ and rotational energy $T$. The important point, however, 
is the fact that the accuracy is improved as the grid number increases. We are thus confident that our new formulation has 
indeed succeeded in finding two-layered configurations in rotational equilibrium. 
\subsection{$j$-constant and $v$-constant cases}
We move on to the combinations of other rotation laws. In the following configurations, each layer is rotating differentially. 
Although various combinations are actually possible, only those in the same rotation law are considered here just for simplicity.
Thus the $j$-constant case refers to the configurations, in which each layer obeys the $j$-constant law in Eq.~(\ref{eq:jconst})
although $h_{0}$ and $A$ are different between the layers. The same is true of the $v$-constant case. For each case 
we have constructed two configurations that have $(r_p,r_{E_{12}},r_{P_{12}},A_{(1)},A_{(2)})=(0.6,0.5,0.32,0.1,0.05)$ and $(0.6,0.5,0.5,0.1,0.1)$. 
Note again that $h_{0(i)}$, or how fast each layer is rotating, are not specified but solved and $r_{E_{12}}$ and $r_{P_{12}}$
are the control parameters. On the other hand, the degree of differential rotations or $A_{(i)}$ can be specified freely.

\begin{figure*}
  \begin{center}
  \vspace*{40pt}
    \begin{tabular}{cc}
      \hspace{-2.0cm}
      \begin{minipage}{0.5\hsize}
      \includegraphics[width=12.0cm]{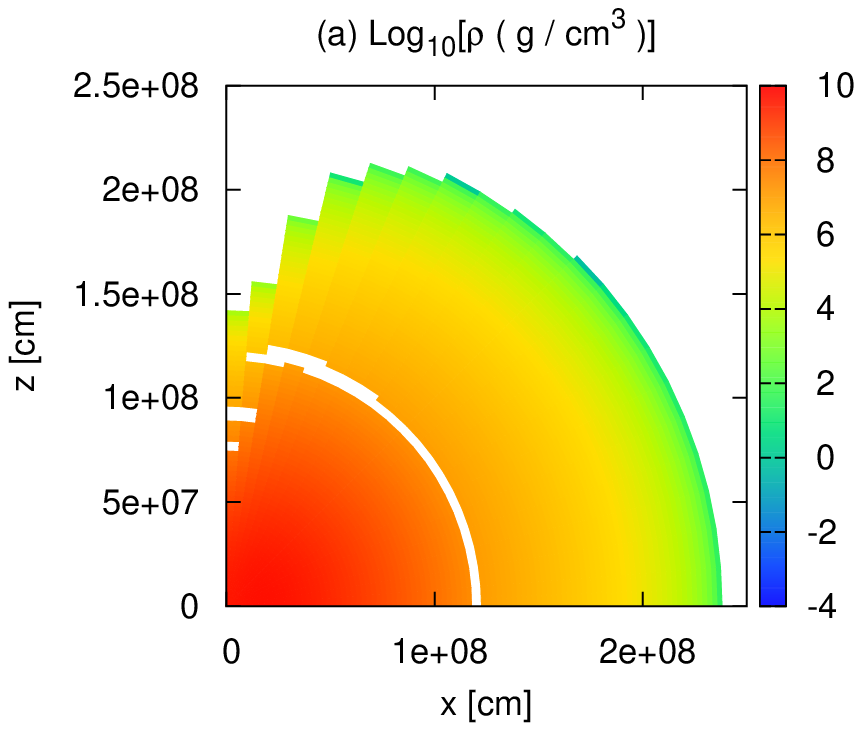}
      \end{minipage}
      \hspace{-1.0cm}
      \begin{minipage}{0.5\hsize}
      \includegraphics[width=12.0cm]{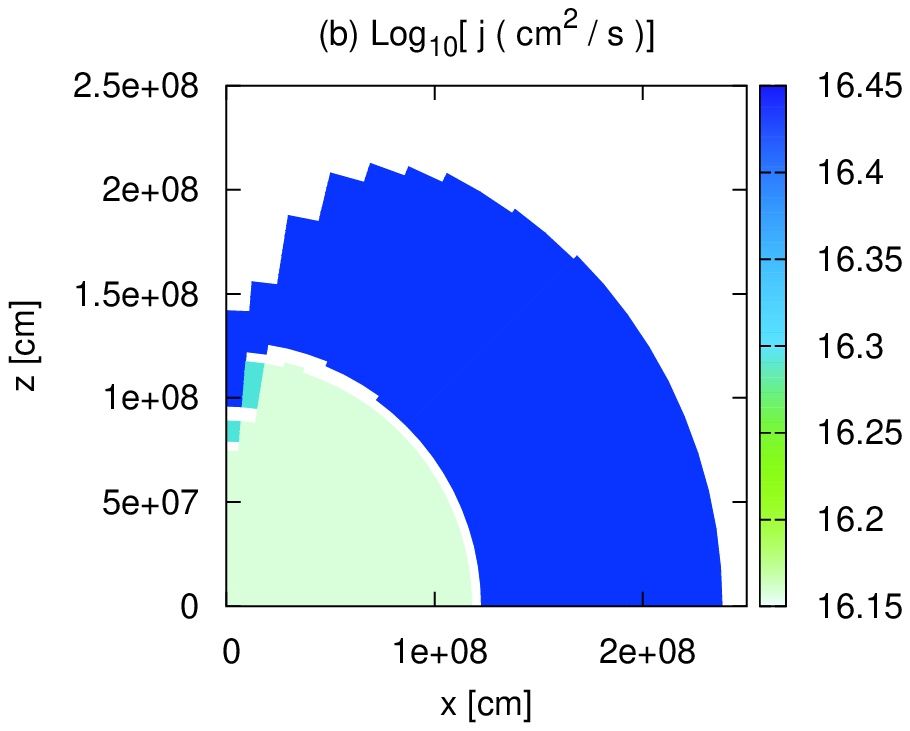}
      \end{minipage}
    \end{tabular}
    \caption{\label{fig:j-const}
    Contour plots in the meridional section of the density (left panel) and angular velocity (right panel) for 
    the $j$-constant case with $(r_p,r_{E_{12}},r_{P_{12}})=(0.6,0.5,0.32),A_{(1)}=0.1$ and $A_{(2)}=0.05$. 
    The contour levels are equally spaced in the logarithmic scale in both panels. 
    The thick curves in the stellar interiors correspond to the layer boundary 
    in both panels.
    }
  \end{center}
\end{figure*}

\begin{figure*}
  \begin{center}
  \vspace*{40pt}
    \begin{tabular}{cc}
      \hspace{-2.0cm}
      \begin{minipage}{0.5\hsize}
      \includegraphics[width=12.0cm]{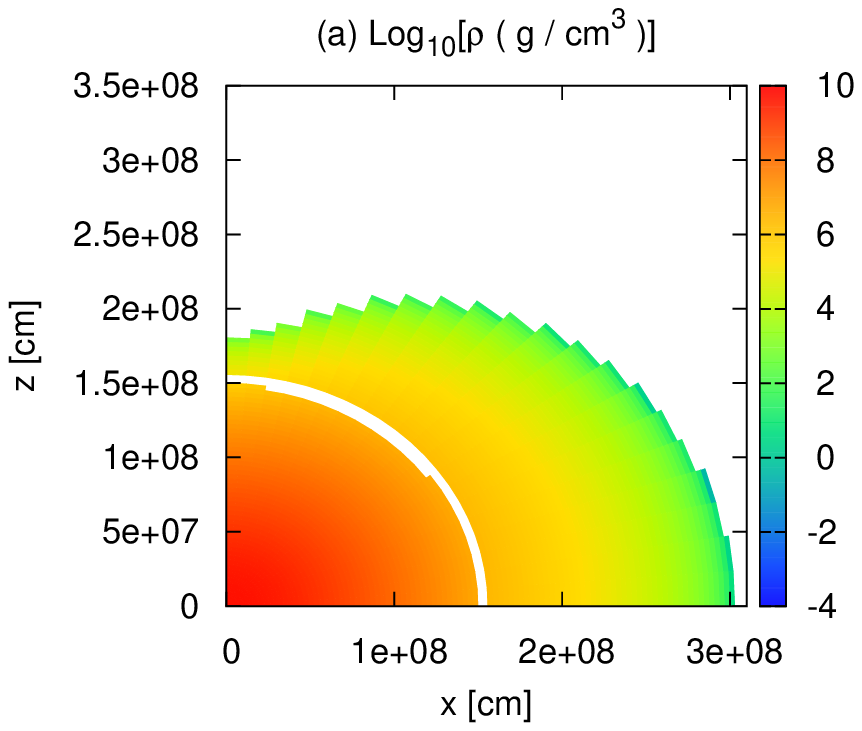}
      \end{minipage}
      \hspace{-1.0cm}
      \begin{minipage}{0.5\hsize}
      \includegraphics[width=12.0cm]{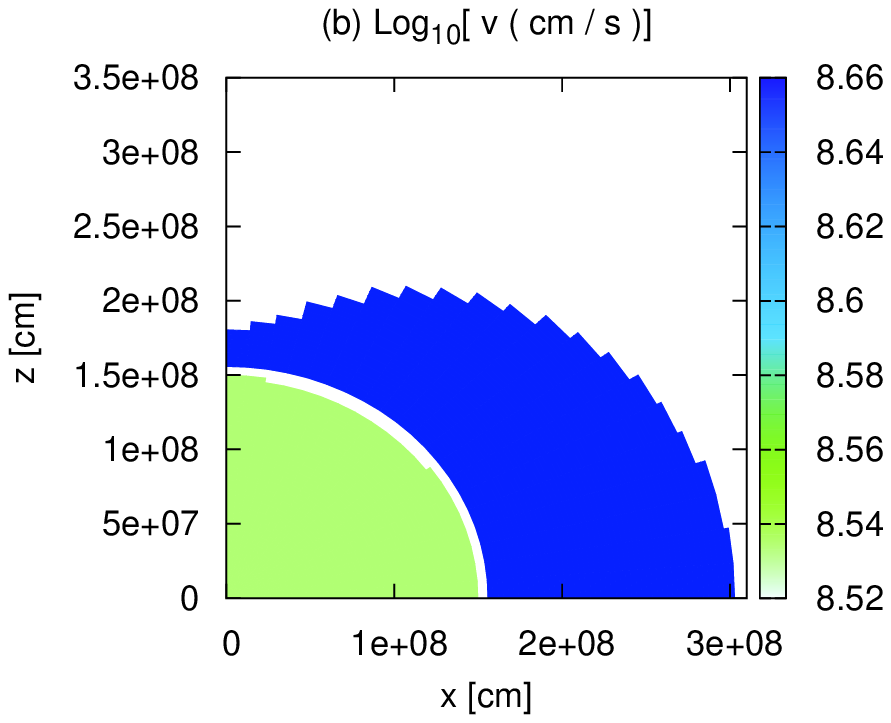}
      \end{minipage}
    \end{tabular}
    \caption{\label{fig:v-const}
    Contour plots in the meridional section of the density (left panel) and angular velocity (right panel) for 
    the $v$-constant case with $(r_p,r_{E_{12}},r_{P_{12}})=(0.6,0.5,0.5),A_{(1)}=0.1$ and $A_{(2)}=0.1$.
    The contour levels are equally spaced in the logarithmic scale in both panels. 
    The thick curves in the stellar interiors correspond to the layer boundary in both panels.
    }
  \end{center}
\end{figure*}

Table~\ref{tab:twolay-D} summarizes the quantities that characterize the configurations whereas Figs.~\ref{fig:j-const} and 
\ref{fig:v-const} display the contour plots of the density and angular velocity for the models in the $j$-constant and 
$v$-constant cases, respectively. As expected and shown in these figures, the differentially rotating models are more 
deformed than the one presented in the previous section, in which each layer rotates rigidly. The values of $VC$ in Table~\ref{tab:twolay-D} are of the same order,
$\sim 10^{-5}$, as in the non-rotational and $\Omega$-constant cases, that fact indicates our formulation's capability of 
constructing configurations with strongly differential rotations. This may be important in dealing with the progenitors of GRB~\citep{Woosley:2005gy}. 

\begin{table}
\centering
\caption{\label{tab:twolay-R} 
Radius, mass, angular velocities, Bernoulli constants 
and virial for two $\Omega$-constant configurations.  
}
\begin{tabular}{cccccccc}
\hline\hline
$(r_p,r_{E_{12}},r_{P_{12}})$ & $R~[{\rm cm}]$ & $M~[M_\odot]$ & 
$\Omega_{(1)}~[{\rm rad} / {\rm s}]$ & $\Omega_{(2)}~[{\rm rad} / {\rm s}]$ & $c_{(1)}$ & $c_{(2)}$ & $VC$\\
\hline
(0.9,0.5,0.5)   & 2.802E+08 & 1.554E+00 & 1.441E+00 & 1.087E+00 & -2.207E-03 & -2.911E-03 & 2.120E-05\\
(0.8,0.5,0.433) & 3.116E+08 & 1.585E+00 & 1.854E+00 & 2.910E+00 & -1.838E-03 & -2.453E-03 & 1.499E-05\\
\hline
\end{tabular}
\end{table}

\begin{table}
\centering
\caption{\label{tab:twolay-D}
Radius, mass, specific angular momenta/rotational velocities, Bernoulli constants, 
and virial for $j$-constant and $v$-constant cases. 
}
\begin{tabular}{cccccccc}
\hline\hline
\multicolumn{8}{c}{j-constant case}\\
$(r_p,r_{E_{12}},r_{P_{12}},A_{(1)},A_{(2)})$ & 
$R~[{\rm cm}]$                                & 
$M~[M_\odot]$                                 & 
$j_{0(1)}~[{\rm cm^2} / {\rm s}]$             & 
$j_{0(2)}~[{\rm cm^2} / {\rm s}]$             & $c_{(1)}$ & $c_{(2)}$ & $VC$\\
\hline
(0.6,0.5,0.32,0.1,0.05) & 2.405E+08 & 1.739E+00 & 2.683E+16 & 1.435E+16 & -3.501E-03 & -4.956E-03 & 4.969E-05\\
(0.6,0.5,0.5,0.1,0.1)   & 2.555E+08 & 1.711E+00 & 2.748E+16 & 2.074E+16 & -2.870E-03 & -4.042E-03 & 4.404E-05\\
\hline
\multicolumn{8}{c}{v-constant case}\\
$(r_p,r_{E_{12}},r_{P_{12}},A_{(1)},A_{(2)})$ & 
$R~[{\rm cm}]$                                & 
$M~[M_\odot]$                                 & 
$v_{0(1)}~[{\rm cm} / {\rm s}]$             & 
$v_{0(2)}~[{\rm cm} / {\rm s}]$             & $c_{(1)}$ & $c_{(2)}$ & $VC$\\
\hline
(0.6,0.5,0.32,0.1,0.05) & 3.037E+08 & 1.823E+00 &  4.694E+08 & 4.709E+08 & -1.837E-03 & -2.434E-03 & 2.204E-05\\
(0.6,0.5,0.5,0.1,0.1)   & 3.057E+08 & 1.654E+00 &  4.506E+08 & 3.412E+08 & -1.632E-03 & -2.295E-03 & 1.326E-05\\
\hline
\end{tabular}
\end{table}

\section{Summary and Discussions}\label{sec:sum}

Bearing in mind the application to the study of rotational massive stars in their late evolutionary phase, in this paper 
we have proposed a new formula to construct multi-layered configurations in rotational equilibrium.
This is an extension of the Hachisu self-consistent field scheme that is based on the Bernoulli equation
and meant originally for single-layered configurations that are rotating cylindrically with a barotropic EOS. 
In our method, on the other hand, each layer is assumed to rotate still cylindrically with a barotropic EOS 
but the rotation laws and EOS's are different from layer to layer. We have shown that the pressure should 
be continuous at the layer boundary whereas the density is in general discontinuous across the boundary, 
which is an alternative demonstration that the EOS cannot be identical for the adjacent layers. We have 
identified the variables that are appropriate to make the iteration scheme convergent. This is indeed a crucial 
ingredient in our formula. 

For demonstration, we have actually constructed several configurations with two layers for three representative 
rotation laws, which we have referred to as the $\Omega$-constant, $j$-constant and $v$-constant laws in this paper. 
We have found that the virial equation is satisfied with a typical error of $10^{-5}$ irrespective of the rotation laws
if we deploy $1000 \times 200$ mesh points and we have also demonstrated that the error is reduced as 
the number of mesh points is increased. Incidentally, it has been confirmed that a non-rotational configuration is 
also reproduced by the present scheme. From these results it is obvious that our method works well and is robust indeed.
The application of the present formula to more realistic problems will be published elsewhere~\citep{Nagakura:2010}. 

As commented in Sec.~\ref{sec:basic}, it is straightforward to extend our scheme to the configurations with 
more than two layers though the procedure becomes a bit more involved. Although we have combined the rotation
laws of the same family but with different parameters for simplicity in this paper, two rotation laws of different families
can be treated in the same way. The implementation of more realistic EOSs will pose no problem in principle as long as they are 
barotropic. 
We may employ the idea by \citet{Jackson:2005,MacGregor:2007zy} that the pressure, density, and temperature are 
assumed to be functions of the effective potential alone.  
Moreover, the present formula will be able to treat configurations with a topology of torus by 
relaxing the assumption that the surface extends itself to the symmetry axis and by choosing appropriate points on the 
equator to impose the conditions corresponding to Eqs.~(\ref{eq:cond4}), (\ref{eq:cond3}) and (\ref{eq:cond6}) although
we do not know how realistic such configurations are.  

In our formula, the layer boundary is determined from the condition that the pressure be continuous there.
In reality, however, the layers in the stellar interior correspond to the regions of different chemical compositions
and their boundaries are determined by the thermodynamical conditions for nuclear burnings. This difference 
originates from the fact that we have imposed piece-wise cylindrical rotation laws. In the actual stellar
interior, each layer obeys a boroclinic EOS and, as a result, rotates non-cylindrically. Moreover, the gas motions in 
the meridian section such as convections and meridional circulations are likely to exist generically. Then
the original partial differential equations should be solved somehow, which is a formidable task and will need an 
entirely new approach. Our formula, therefore, is admittedly a rather crude approximation to the reality but, 
hopefully, not so bad one if one chooses an appropriate rotation law for each layer. In fact, it will be much better than
any approximate configurations with only a single-layer.

The real challenge will be to somehow implement chemical evolutions to the sequence of rotational
configurations. One possibility may be an extension of the idea employed in most of the current one dimensional
evolution models of rotational massive stars~\citep{Heger:2000ud,Hirschi:2004ks,Limongi:2000km}. Under the 
assumption that the thermodynamical conditions as well as the chemical abundances are uniform  on each 
surface of constant effective potential, we solve the nuclear network locally and then transfer generated energy 
on the multi-dimensional mesh. The transport of angular momentum may be also approximated by diffusion. Since 
the resultant distributions of thermodynamical quantities and elements will in general be non-uniform on the surface of
constant effective potential, we will take their angular averages on the surface and solve the new rotational 
equilibrium for the obtained equations of state and rotation law. This completes the single cycle and the iteration of
this process will give the temporal evolution of rotational stars. 
We hope that this procedure is feasible and that the formulation presented in this paper will contribute to the study of the 
influences of non-sphericity on the evolution of rapidly rotating massive stars.

\acknowledgments
Numerical computations were in part carried on XT4 and 
general common use computer system at the center for Computational 
Astrophysics, CfCA, the National Astronomical Observatory of Japan 
and on NEC-SX8 at Yukawa Institute for Theoretical Physics in Kyoto University.
This study was supported in part by the Grants-in-Aid for the Scientific Research 
from the Ministry of Education, Science and Culture of Japan (Nos. 80251403 and 19104006).


\end{document}